\documentclass[english, letterpaper, 10 pt, conference]{ieeeconf}
\usepackage[T1]{fontenc}
\usepackage[latin9]{inputenc}
\usepackage{amsmath}
\usepackage{cite}
\usepackage[caption=false]{subfig}
\usepackage{dsfont}
\usepackage{bm}
\usepackage{url}

\IEEEoverridecommandlockouts
\makeatletter

\usepackage{tikz}
\usepackage{tikzscale}
\usepackage{amssymb}

\usepackage{enumerate}   
\usepackage{graphicx} 
\usepackage[binary-units=true]{siunitx}
\usepackage{pgfplots}
\pgfplotsset{compat = 1.12}
\usepackage{cleveref}
\usepackage[ruled,vlined]{algorithm2e}
\usepackage{booktabs}
\usetikzlibrary{external}
\pgfplotsset{select coords between index/.style 2 args={
		x filter/.code={
			\ifnum\coordindex<#1\fi
			\ifnum\coordindex>#2\fi
		}
}}
\usepgfplotslibrary{groupplots}
\usepgfplotslibrary{fillbetween}
\usetikzlibrary{matrix}

\newcounter{tagnumb}
\newcommand{\bx}{\xi}
\newcommand{\bu}{\mathbf u}

\crefname{algocf}{alg.}{algs.}
\Crefname{algocf}{Algorithm}{Algorithms}

\newcommand{\includegtikz}[2][]{ 
	\tikzsetnextfilename{#2}
	\includegraphics[#1]{figures/tikz/#2} }

\makeatother

\definecolor{col1}{rgb:Hsb}{210,1,1}
\definecolor{col2}{rgb:Hsb}{220,1,0.8}
\definecolor{col3}{rgb:Hsb}{230,1,0.6}
\definecolor{col4}{rgb:Hsb}{240,1,0.4}
\definecolor{col5}{rgb:Hsb}{250,1,0.2}

\usepackage{babel}

\title{\LARGE \bf High-Speed Trajectory Planning for Autonomous Vehicles Using a Simple Dynamic Model}

\author{Florent Altch\'e$^{2,1}$, Philip Polack$^{1}$ and Arnaud de La Fortelle$^{1}$
\thanks{$^{1}$ MINES ParisTech, PSL Research University, Centre for robotics, 60 Bd St Michel 75006 Paris, France {\tt\small [florent.altche, philip.polack, arnaud.de\_la\_fortelle] @mines-paristech.fr}}
\thanks{$^{2}$ \'Ecole des Ponts ParisTech, Cit\'e Descartes, 6-8 Av Blaise Pascal, 77455 Champs-sur-Marne, France}%
}

\begin{document}

\maketitle
\thispagestyle{empty}
\pagestyle{empty}

\begin{abstract} 
	To improve safety and energy efficiency, autonomous vehicles are expected to drive smoothly in most situations, while maintaining their velocity below a predetermined speed limit. However, some scenarios such as low road adherence or inadequate speed limit may require vehicles to automatically adapt their velocity without external input, while nearing the limits of their dynamic capacities. Many of the existing trajectory planning approaches are incapable of making such adjustments, since they assume a feasible velocity reference is given. Moreover, near-limits trajectory planning often implies high-complexity dynamic vehicle models, making computations difficult. In this article, we use a simple dynamic model derived from numerical simulations to design a trajectory planner for high-speed driving of an autonomous vehicle based on model predictive control. Unlike existing techniques, our formulation includes the selection of a feasible velocity to track a predetermined path while avoiding obstacles. Simulation results on a highly precise vehicle model show that our approach can be used in real-time to provide feasible trajectories that can be tracked using a simple control architecture. Moreover, the use of our simplified model makes the planner more robust and yields better trajectories compared to  kinematic models commonly used in trajectory planning.
\end{abstract}

\section{Introduction}
In order to improve safety and energy efficiency and not to surprise other road users, autonomous vehicles are widely expected to drive smoothly; moreover, traffic efficiency concerns will likely lead these vehicles to follow relatively closely the road speed limit. However, such speed limits may not always exist (\textit{e.g.}, on some dirt roads) or they may not be followed safely due to the road topology (\textit{e.g.}, mountain roads with sharp curves) or weather conditions (\textit{e.g.}, wet or icy road).

Many trajectory planning algorithms rely on an \textit{a-priori} knowledge of a target velocity that can either be an explicit parameter of the problem~\cite{Benenson2008,Qian2016} or be implicitly given by requiring a set of target positions at fixed times~\cite{Falcone2008,Frasch2013}. Some authors have proposed using the road curvature~\cite{koh2016speed} to provide an upper bound on the velocity, which corresponds to a maximum lateral acceleration; this method can be extended to obstacle avoidance by first planning a collision-free path, then adjusting the velocity consequently~\cite{Kim2011}. However, as path selection and velocity planning are intrinsically linked, this approach can lead to severe inefficiencies.

For this reason, model predictive control (MPC) techniques are often used in the literature, since they allow to simultaneously compute a feasible trajectory and a sequence of control inputs to track it. However, high-speed trajectory planning requires a complex modeling of the vehicle to account for its dynamic limitations, which mainly come from the complex and highly non-linear~\cite{Pacejka2005} tire-road interactions. Adding to this difficulty, wheel dynamics are generally much faster (around \SI{1}{\milli\second}~\cite{Altrock1994}) than changes of the vehicle's macroscopic state (typically \SI{100}{\milli\second}). Therefore, the existing literature is generally divided between medium-term (a few seconds) trajectory planning including obstacle avoidance for low-speed applications, mainly relying on simple kinematic models (see, \textit{e.g.}~\cite{Abbas2014,Cardoso2016}), and short-term (sub-second) trajectory tracking for high-speed or low-adherence applications using wheel dynamics modeling (see, \textit{e.g.},~\cite{Falcone2007,Park2009,Gao2010,liniger2015optimization}). In the second case, obstacle avoidance is generally not considered (with the notable exception of~\cite{Liu2014}), and the existence of a feasible collision-free trajectory is not guaranteed in the case of an unexpected obstacle. Note that sampling-based approaches~\cite{JeonghwanJeon2013} have also been proposed in the literature; however, they do not provide optimality guarantees when a finite number of samples is selected, and may have trouble finding a feasible solution in complex scenarios.

In this article, we propose a middle-ground approach to allow trajectory planning over a few seconds in high dynamic situations. This approach relies on the use a simple vehicle model (initially proposed in~\cite{Altche2017}) derived from a realistic dynamic modeling of the vehicle body and wheels, which accounts for tire slip effects using carefully estimated bounds on the dynamics. Using this model, we design a non-linear trajectory planning framework, which is able to follow a predefined path (\textit{e.g.}, the road centerline) at high speed while avoiding obstacles. Comparison with a more classical kinematic bicycle model shows that our proposed planner provides better trajectories and is more robust. Moreover, our simulations showed that computation time remains below \SI{80}{\milli\second} on a standard computer and could be further reduced using auto-generation~\cite{Frasch2013}, thus allowing real-time use. To the best of the authors' knowledge, no comparable framework has been published for medium- or long-term planning in high dynamic situations.

The rest of this article is structured as follows: in \Cref{sec:vehicle-model}, we present a 9 degrees of freedom dynamic model of the vehicle's body, that we use throughout the rest of this article. In \Cref{sec:mpc}, we briefly describe our simplified dynamic model and to formulate a trajectory planner based on Model Predictive Control, which computes an aggressive yet feasible trajectory. This planner is validated and its performance is compared to a classical kinematic bicycle model in \Cref{sec:simulations} using a realistic physics simulation suite; finally, \Cref{sec:conclusion} concludes the study.

\section{Vehicle model\label{sec:vehicle-model}}
\begin{figure}[h!]
	\centering
	\includegraphics[scale=0.4]{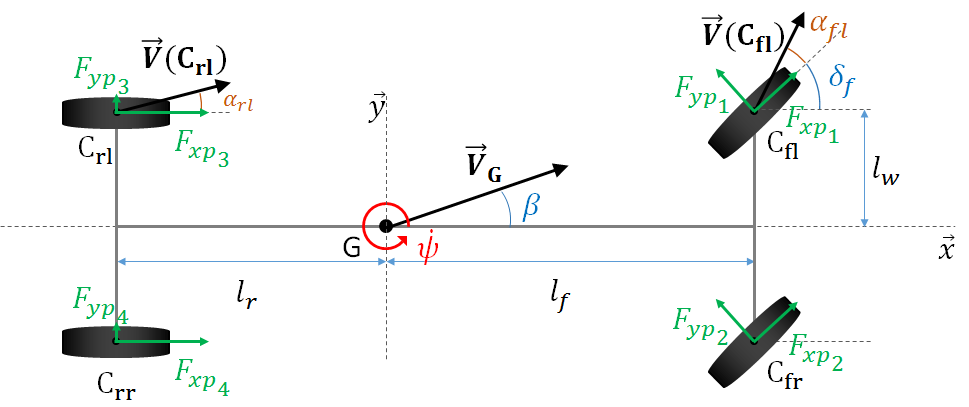}       
	\caption{Simulation model of the vehicle in the $(x,y)$ plane}
	\label{fig:carSim}
\end{figure}

In this article, we consider a 9 degrees of freedom modeling of the vehicle body, which provides a satisfying balance between accuracy and computational cost. Alongside with the usual 2D state $[X, Y, \psi]$ (with $\psi$ the yaw rotation) of the vehicle, the model takes into account roll and pitch movements, wheel dynamics and coupling of longitudinal and lateral tire slips. Being a chassis model, it does not take into account the dynamics of the car engine or brakes. The control inputs of the vehicle are the torque $T_i$ applied to each wheel $i$ and the steering angle of the front wheels, $\delta$. We use uppercase letters (\textit{e.g.}, $X$, $Y$) to denote coordinates in the ground (global) frame, and lowercase letters for coordinates in the vehicle (local) frame; the $x$ coordinate in the local frame corresponds to the longitudinal component. The notations are given in \Cref{tab:notations} and illustrated in \Cref{fig:carSim}.


\begin{table}[h]
	\caption{Notations}
	\label{tab:notations}
	\begin{tabular}{p{1.45cm} p{6.35cm}}
		\toprule
		$X$, $Y$, $Z$ & Position of the vehicle's CoM (ground frame) \\
		$\theta$, $\phi$, $\psi$ & Roll, pitch and yaw angles of the car body \\
		$V_x$, $V_y$ & Longitudinal and lat. vehicle speed (vehicle frame) \\
		$V_{xw_i}$ & Longitudinal speed of wheel $i$ (wheel frame)\\
		$\omega_i$ & Angular velocity of wheel $i$ \\
		$\zeta_i$ & Displacement of suspension $i$ \\
		$\delta$ & Steering angle of the front wheels \\
		$T_{\omega_i}$ & Total torque applied to wheel $i$\\
		$F_{xw_i}$, $F_{yw_i}$  & Longitudinal and lateral forces on wheel $i$ (wheel frame)\\
		$F_{x_i}$, $F_{y_i}$ & Longitudinal and lat. forces on wheel $i$ (vehicle frame)\\
		$F_{z_i}$ & Normal ground force on wheel $i$\\
		$F_{aero}$ & Air drag force on the vehicle \\
		$M_T$ & Total mass of the vehicle\\
		$I_x$, $I_y$, $I_z$ & Roll, pitch and yaw inertia of the vehicle \\
		$I_{r_i}$ & Inertia of wheel $i$ around its axis \\
		$l_f$, $l_r$ & Distance between the front/rear axle and the CoM\\
		$l_w$ & Half-track of the vehicle\\ 
		$r_{w}$ & Effective radius of the wheels \\ 
		$k_s$, $d_s$ & Suspensions stiffness and damping \\
		\bottomrule
	\end{tabular}
\end{table}

We make the assumptions that the body of the vehicle rotates around its center of mass, and that the aerodynamic forces do not create a moment on the vehicle. Moreover, we assume that the road remains horizontal, and any slope or banking angle is neglected; this assumption could be relaxed using a slightly more complex vehicle model. Under these hypotheses, the dynamics of the vehicle's center of mass are written as:
\begin{subequations}\label{eq:globalframe}
	\begin{align}
	\dot{X} =\ & V_x \cos\psi - V_y \sin\psi\\
	\dot{Y} =\ & V_x \sin\psi + V_y \cos\psi	\\
	\dot{V}_x =\ & \dot{\psi} V_y + \frac{1}{M_T}\sum_{i=1}^4 F_{x_i} - F_{aero}\\
	\dot{V}_y =\ & - \dot{\psi} V_x + \frac{1}{M_T} \sum_{i=1}^4 F_{y_i},
	\end{align}
\end{subequations}
where $F_{x_i}$ and $F_{y_i}$ are respectively the longitudinal and lateral tire forces generated on wheel $i$, expressed in the local vehicle frame $(x,y)$. The yaw, roll and pitch motions of the car body are computed as:
\begin{subequations}\label{eq:rollpitch}
\begin{align}
	I_z\ddot{\psi} = l_f (F_{y_1} + F_{y_2})  - l_r (F_{y_3} + F_{y_4}) &  \nonumber\\
	 +\, l_w (F_{x_2}+F_{x_4}-\, & F_{x_1}-F_{x_3}) \\
	I_x\ddot{\theta} = l_w (F_{z_1}+F_{z_3}-F_{z_2}-F_{z_4})\, +\, & Z \sum_{i=1}^4 F_{y_i}\\
	I_y\ddot{\phi} = l_r (F_{z_3} + F_{z_4}) -l_f (F_{z_1} +F_{z_2})\, -\, & Z \sum_{i=1}^4 F_{x_i}
\end{align}
\end{subequations}
where $F_{z_i} = - k_s \zeta_{i}(\theta, \phi) - d_s \dot{(\zeta_{i})}(\theta, \phi)$, with $\zeta_i(\theta,\phi)$ the displacement of suspension $i$ for the given roll and pitch angles of the car body. The variation of $F_z$ models the impact of load transfer between tires. Finally, the dynamics of each wheel $i$ can be written as
\begin{align} 
	I_r \dot{\omega}_i = T_{{\omega}_i}-r_w F_{xw_i}.
\end{align}

In general, the longitudinal and lateral forces $F_{xw_i}$ and $F_{yw_i}$ depend on the longitudinal slip ratio $\tau_{i}$, the side-slip angle $\alpha_i$, the reactive normal force $F_{z_i}$ and the road friction coefficient $\mu$. The slip ratio of wheel $i$ can be computed as
\begin{equation}\label{eq:slip-ratio}
	\tau_i = \left\{ 
		\begin{array}{l l}
			\frac{r_w \omega_i - V_{xw_i}}{r_w\omega_i} & \text{if } r_w\omega_i \geq V_{xw_i} \\
			\frac{r_w \omega_i - V_{xw_i}}{V_{xw_i}} & \text{otherwise}.
		\end{array} 
		\right.
\end{equation}

The lateral slip-angle $\alpha_i$ of tire $i$ is the angle between the wheel's orientation and its velocity, and can be expressed as
\begin{align}
\alpha_f &= \delta - \frac{V_y + l_f \dot{\psi}}{V_x \pm l_w \dot{\psi}} \\
\alpha_r &= - \frac{V_y - l_r \dot{\psi}}{V_x \pm l_w \dot{\psi}}
\end{align}
where $f$ and $r$ denote the front and rear wheels respectively.

In this article, we use Pacejka's combined slip tire model (equations (4.E1) to (4.E67) in~\cite{Pacejka2005}), which takes into account the interaction between longitudinal and lateral slips, 
thus encompassing the notion of friction circle~\cite{guntur1980friction}. For clarity purposes, we do not reproduce the complete set of equations here.



%
%

\section{Constrained second-order integrator model\label{sec:2d-integrator}}
Theoretically, it is possible to use the 9 degrees of freedom model inside a model predictive control framework to directly compute an optimal trajectory and the corresponding controls. However, classical optimization tools often struggle to solve problems involving highly nonlinear constraints or cost functions, as it is the case in the model presented in \Cref{sec:vehicle-model} -- notably due to disjunction~\eqref{eq:slip-ratio} which makes $\tau$ non-differentiable. Additionally, wheel dynamics generally occur over very small characteristic times -- typically a few milliseconds -- which requires choosing a correspondingly small discretization time step, making planning over long horizons impractical at best. For this reason, simplified models are very often preferred. Kinematic bicycle (or single-track) models~\cite{Kong2015,Abbas2014,Cardoso2016}, or even simpler second-order integrator models~\cite{Qian2016} are therefore common in the trajectory planning literature. 

One of the main issues of these simplified models is that they are generally considered to be imprecise when nearing the handling limits of the vehicle. To counter this problem, previous work~\cite{Altche2017} 
presented a constrained second-order dynamic model derived from simulation data using the 9 degrees of freedom model of \Cref{sec:vehicle-model}. For lack of space, we only briefly described these previous results here. 

\begin{figure}[h!]
	\subfloat[Lateral vs. longitudinal\label{fig:feasible-axay-fullvy}]{
		\includegtikz[width=0.9\columnwidth,height=5cm]{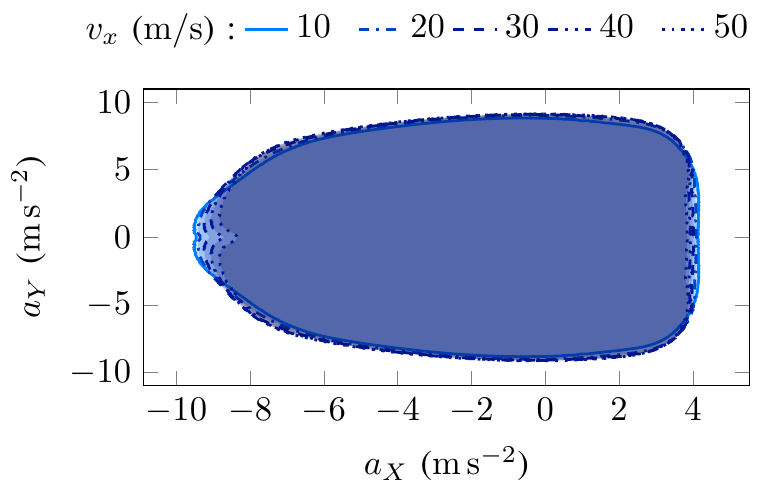}
	} 
		
	\subfloat[Yaw vs. lateral\label{fig:feasible-ayaY-fullvy}]{
		\includegtikz[width=0.9\columnwidth,height=4cm]{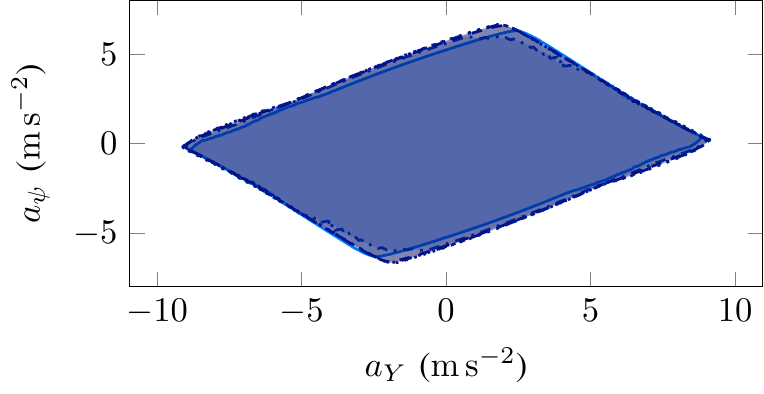}
	}
	\caption{Envelope of the sets of feasible accelerations for different longitudinal velocities $v_{x,0}$ and lateral velocity $v_{y,0} \in [-0.2 v_{x,0}, 0.2 v_{x,0}]$; notice the slight deformation along the $a_X$ axis with increasing $v_{x,0}$.\label{fig:feasible}.}
\end{figure}

Using a random sampling method, we first compute (off-line) an envelope for the set of feasible longitudinal ($a_X$), lateral ($a_Y$) and angular ($a_\psi$) accelerations, as shown in \Cref{fig:feasible}. More specifically, these sets are computed for a given longitudinal velocity $v_{x,0}$ and a lateral velocity $v_{y,0} \in [-0.2 v_{x,0}, 0.2 v_{x,0}]$. Therefore, only part of this region is actually reachable from an initial vehicle state (corresponding to a determined lateral velocity). In this article, we assume that the lateral dynamics of the vehicle is sufficiently fast to neglect this effect. Simulation results (see \Cref{sec:simulations}) show this assumption seems reasonable. 

Using these results, we propose a constrained double integrator model for the vehicle dynamics. This model considers a state vector $\bx = [X, Y, \psi, v_x, v_y, v_\psi]^T$ and a control $\bu = [u_x, u_y, u_\psi]^T$, with the same notations and reference frames as presented in \Cref{sec:vehicle-model}. The dynamic equation of the system is $\dot \bx = f_{2di}(\bx, \bu)$
with \begin{equation}f_{2di}\left(\bx, \bu \right) = \left[ \begin{array}{c}v_x \cos \psi - v_y\sin \psi \\  v_x \sin \psi +v_y \cos \psi \\ {[}v_\psi, u_x, u_y, u_\psi{]^T} \end{array} \right].\end{equation}

\begin{figure}[h!]
	\centering \subfloat[Feasible set (blue) and approximation (red)]{\includegtikz[width=0.61\columnwidth,height=2.5cm]{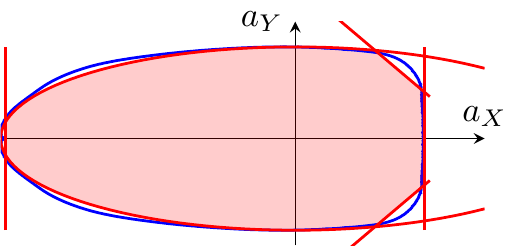}} \subfloat[Detail]{\includegtikz[width=0.3\columnwidth,height=2.5cm]{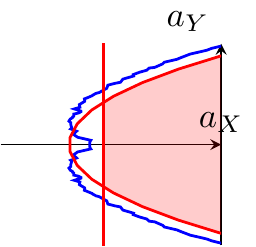}}
	\caption{Convex approximation of the feasible accelerations in the $(a_X, a_Y)$ plane. The actual region is shown in blue. \label{fig:polygon-approx}}
\end{figure}

\begin{figure}
	\includegtikz[width=0.9\columnwidth,height=3cm]{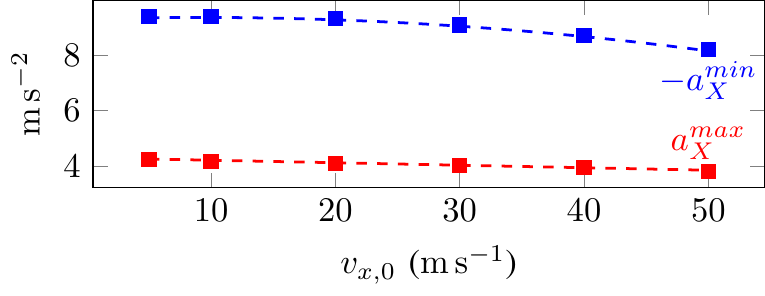}
	\caption{Variations of the $a_X^{min}$ and $a_X^{max}$ coefficients with the initial longitudinal velocity $v_{x,0}$; the dashed lines show the polynomial fit. \label{fig:alpha-beta}}
\end{figure}

To allow the use of this model in planning, we approximate the sets shown in \Cref{fig:feasible} as a set of convex linear and nonlinear constraints in the $(a_X, a_Y)$ plane as shown in \Cref{fig:polygon-approx}. These constraints are expressed as: \begin{align}
\left( \frac{a_X}{\alpha} \right)^2 + \left( \frac{a_Y}{\beta} \right)^2 \leq 1 \\
a_X^{min}(v_{x,0}) \leq a_X  \leq a_X^{max}(v_{x,0}) \\
A [a_X, a_Y, a_\psi]^T  \leq b
\end{align}
where $A$ is a constant matrix, $b$ a constant vector and $a_X^{min}$, $a_Y^{min}$ depend on $v_{x,0}$. For our model vehicle, the experimental data of \Cref{fig:feasible} yield $\alpha =$ \SI{9.4}{\meter \per \second\squared},  $\beta =$ \SI{9.0}{\meter \per \second\squared}, $A = \left( \begin{smallmatrix} 2.6 & 1 \\ 2.6 & -1 \end{smallmatrix} \right)$ and $b = \left( \begin{smallmatrix} 15.3 \\ 15.3 \end{smallmatrix} \right)$ \SI{}{\meter\per\second\squared}. \Cref{fig:alpha-beta} shows the variation of $a_X^{min}$ and $a_X^{max}$ with the initial longitudinal velocity; a polynomial fit yields $a_X^{min}(v_{x,0}) = -9.3 - 0.013v_{x,0} + 0.00072{v_{x,0}}^2$ and $a_X^{max}(v_{x,0}) = 4.3 - 0.009v_{x,0}$ (with $v_{x,0}$ expressed in \SI{}{\meter\per\second}). In the $(a_Y, a_\psi)$ plane, we suppose a linear relation between $a_\psi$ and $a_Y$ in the form: $a_\psi = \gamma a_Y$, with $\gamma =$ \SI{0.56}{\radian \per \meter}. Although seemingly restrictive, this approximation provides better results than using a parallelogram acceptable region, and in practice corresponds to minimizing the vehicle slip angle.

\section{MPC formulation for trajectory planning\label{sec:mpc}}
We now use the second-order integrator model developed in the previous section to design a trajectory planner based on model predictive control. In this section, we assume that the vehicle tries to follow a known reference path, for instance the centerline of a given lane. The reference path $\gamma_{ref}$ is supposed to be given as a set of positions $(X_{ref}, Y_{ref})$.

In this article, we assume that the vehicle has no information on a safe choice of longitudinal speed. Such situations can arise in various scenarios, such as poor adherence conditions where the speed limit cannot be safely followed, or in the absence of speed limits, for instance on dirt roads or for racing. Moreover, we assume that the vehicle also needs to avoid obstacles on the road; for now, we only consider fixed obstacles with known positions and shapes. Additionally, we do not consider varying road adherence, and we suppose that the tire-road friction coefficient $\mu$ is constant and equal for all four wheels. Future work will study real-time estimation of this friction coefficient to adapted the planner accordingly.

\begin{algorithm}  \DontPrintSemicolon
	\caption{Planning and control\label{alg:control}}
	\KwData{current state $\bx(t_0)$, $\gamma_{ref}$, horizon $T$}
	find $x_{closest}$ := point of $\gamma_{ref}$ closest to $(X(t_0), Y(t_0))$\;
	set $s_0$ := curvilinear position of $x_{closest}$\;
	set $hz$ := $[s_0, s_0+v_x(t_0)T]$\;
	set $p_X$ := fitpolynom($\left.X_{ref}\right|_{hz}$, $s - s_0$, 5)\;
	set $p_Y$ := fitpolynom($\left.Y_{ref}\right|_{hz}$, $s - s_0$, 5)\;
	find $obs$ := list of relevant obstacles\;
	find $\kappa$ := max(abs(curvature($p_X$, $p_Y$, $hz$)))\;
	set $v_{max}$ := min $\left(v_x(t_0) + a_X(B)T, \sqrt{\mu g/\kappa} \right)$\;
	compute $\bx_{mpc}$ := MPC($\bx(t_0)$, $p_X$, $p_Y$, $v_{max}$, $T$, $obs$)\;
	apply low level control to track $\bx_{mpc}$\;
\end{algorithm}

\begin{figure}
	\centering\includegtikz[height=3cm]{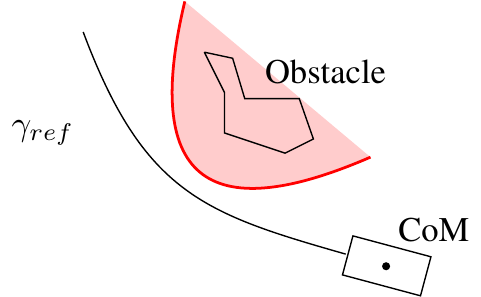}
	\caption{Modeling of an obstacle a parabola. The vertex and roots of the parabola are chosen with enough margin to ensure that no collision can occur as long as the center of mass is outside of the shaded region containing the obstacle. \label{fig:parabola}}
\end{figure}

The planning and control scheme works in several steps, as presented in \Cref{alg:control}. We first approximate  $X_{ref}$ and $Y_{ref}$ over the next planning horizon $T$ as fifth order polynomials of the curvilinear position, starting at the point of $\gamma_{ref}$ closest to the vehicle's current position. Using these polynomials, we compute the maximum (in absolute value) of the path curvature over the planning horizon, noted $\kappa$, to determine an upper bound $v_{max} = \sqrt{\mu g/{\kappa}}$ for the speed of the vehicle in order to limit the lateral acceleration to $\mu g$ (as proposed, \textit{e.g.}, in~\cite{koh2016speed}). Only the relevant obstacles, \textit{i.e.} those for which a risk of collision exists during the next planning horizon $T$, are effectively considered for collision avoidance; we note $\mathcal O$ the set of these obstacles. Generalizing the ideas of~\cite{qian2016hierarchical}, we determine a bounding parabola for each obstacle $o \in \mathcal O$ as shown in \Cref{fig:parabola}, such that the collision avoidance constraints can be written as $p_o(X,Y) \leq 0$, with $p_o$ a second-order multinomial function. Note that the use of unbounded parabolas is preferred to bounded shapes such as ellipses, since they do not create local minima.

At a time $t_0$ corresponding to a vehicle state $\bx(t_0)$ (with an initial longitudinal speed $v_0$), we formulate the motion planning problem using model predictive control with time step duration $h$ and horizon $T = Kh$ as follows:
\begin{subequations}\label{eq:mpc}
	\begin{align}
	\min_{(\bu_k)_{k = 0 \dots K-1}} \ \ & J \big((\bx_k)_{k = 0 \dots K}, (\bu_k)_{k = 0 \dots K-1} \big) \\
	\text{subj. to}\ \  & \bx_{k+1} - \bx_k = h\, f_{2di}(\bx_k, \bu_k) \\
	& \bx_0 = \bx(t_0) \\
	&\left( \frac{u_X}{\alpha} \right)^2 + \left( \frac{u_Y}{\beta} \right)^2 \leq 1 \\
	&a_X^{min}(v_{x,0}) \leq a_X  \leq a_X^{max}(v_{x,0}) \\
	&A [u_X, u_Y]^T  \leq b \\
	&u_\psi = \gamma u_Y \\
	& \forall o \in \mathcal O, \ \ p_o(X_k, Y_k) \leq 0 \label{eq:coll-avoidance} \\
	\text{for}\ \ & k = 0 \dots K-1. \nonumber
	\end{align}
\end{subequations}
Note that, as it is often the case in the planning literature (see, \textit{e.g.}, \cite{Gray2013}), collision avoidance (eq.~\eqref{eq:coll-avoidance}) is actually implemented as soft constraints to avoid infeasibility caused by numerical errors. Additionally, note that our formulation can be slightly modified in order to take moving obstacles into account, by using a different parabola $p_o^k$ for each obstacle and at each time step.

In this article, we only focus on minimizing the deviation from the reference trajectory. In most of the existing MPC literature where a reference speed is supposed to be known in advance, the cost function $J$ is expressed as:
\begin{equation}
J \big((\bx_k), (\bu_k) \big) = \sum_{k=0}^K (X_k - X_k^{ref})^2 + (Y_k - Y_k^{ref})^2.
\end{equation}
In these formulations, $X_k^{ref}$ and $Y_k^{ref}$ implicitly encode the speed at which the reference path should be followed. Since we do not assume that a reference speed is known in advance, this method cannot be applied directly. A possible way to handle this difficulty is to express $Y_k^{ref}$ as a function of $X_k$ (see, \textit{e.g},~\cite{Borrelli2005}). However, this method cannot be applied to all shapes of reference paths, and is notably not suited to sharp turns even when using local instead of global coordinates. For this reason, we introduce an auxiliary state $s$ to denote the curvilinear position of the vehicle along $\gamma_{ref}$, so that $X_k^{ref} =p_X(s_k)$ and $Y_k^{ref} = p_Y(s_k)$. In this article, we use a simple first-order integrator dynamic for $s$ with $\dot s = \sqrt{v_x^2 + v_y^2}$. Noting $\bx'$ the extended state of the vehicle, we instead use the objective function:
\begin{equation}
J = \sum_{k=0}^{K-1} w_v {v_{tol}}_k^2 + w_X {X_{tol}}_k^2 + w_Y {Y_{tol}}_k^2 + w_o {O_{tol}}_k^2
\end{equation}
where $w_v$, $w_X$, $w_Y$ and $w_o$ are positive weighting terms, and we add the following constraints to problem \eqref{eq:mpc}:
\begin{subequations}\label{eq:constr-tol}
	\begin{align}
	{v_{tol}}_k &\ \geq |v_{max} - {v_X}_k| \\
	{X_{tol}}_k &\ \geq |X_k - p_X(s_k)| \\
	{Y_{tol}}_k &\ \geq |Y_k - p_Y(s_k)| \\
	\forall o \in \mathcal O, \ \ {O_{tol}}_k &\ \geq p_o(X_k, Y_k) \\
	&\text{for } k = 0 \dots K-1. \nonumber
	\end{align}
\end{subequations}

\section{Simulation results\label{sec:simulations}}
We used the realistic physics simulator PreScan~\cite{prescan} to validate the proposed MPC trajectory planner. The simulator uses the 9 degrees of freedom model presented in \Cref{sec:vehicle-model}, with the same parameters that were used to obtain the sets of feasible accelerations in \Cref{sec:2d-integrator}. Robustness of the planner to variations of the vehicle parameters is a subject for further study.

The MPC problem \eqref{eq:mpc}-\eqref{eq:constr-tol} is solved using the ACADO Toolkit~\cite{Houska2011}; due to the inner workings of the simulator, the simulation is paused during resolution. The output of the solver is the set of future target longitudinal and lateral accelerations, as well as target future positions and longitudinal velocities for the vehicle in the horizon $T$. These outputs are fed into two low-level controllers, one being tasked with velocity tracking and the other with steering.

Low-level tracking of the planned trajectory is achieved using PID controllers; the lateral control also uses a $\tau$ seconds look-ahead~\cite{Kosecka1997}. At time $t$, the predicted position of the vehicle at $t + \tau$ is computed as $\hat X = X + \tau v_x \cos(\hat \psi)$ and $\hat Y = Y + \tau v_x \cos(\hat \psi)$, with $\hat \psi = \psi + \frac 1 2 \tau \dot \psi$. Instead of tracking the target position at time $t$, the lateral control uses the error between predicted and desired positions at time $t + \tau$. This method was found to provide better performance and stability than a simple PID in our simulations, with a look-ahead duration of $\tau = \SI{0.2}{\second}$. The lateral control takes into account a limited angular velocity for the steering wheel of \SI{12}{\radian\per\second}, which is in the average of recorded steering velocities for human drivers in obstacle avoidance scenarios~\cite{breuer1998}. However, we do not consider the dynamics of the engine or brakes, which are supposed to respond instantaneously.

\begin{figure}
	\includegtikz[width=0.9\columnwidth,height=4cm]{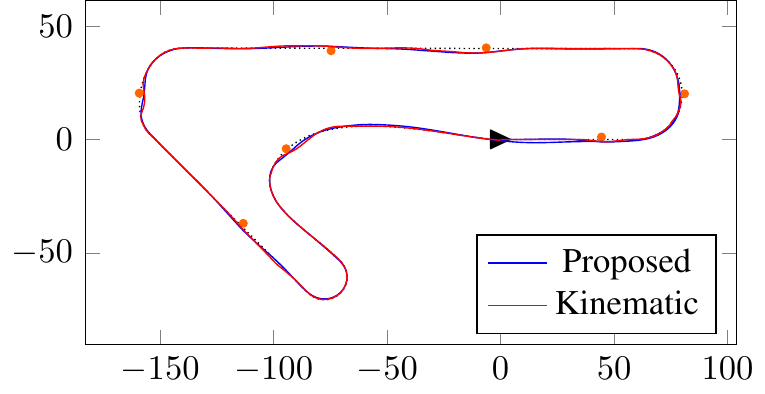}
	\caption{Detail of the reference path (black, dotted) and actual path followed by the vehicle while avoiding obstacles (represented as orange circles), using both planners. 
		\label{fig:output-obst}}
\end{figure}

The reference path used in our simulations is presented in \Cref{fig:output-obst}; the path consists of a \SI{60}{\meter} straight line, a half circle with radius \SI{20}{\meter}, a \SI{200}{\meter} straight line, a Bezier arc corresponding to a 135 degrees turn, a \SI{100}{\meter} straight line, a half circle with radius \SI{10}{\meter} and a final -135 degrees turn Bezier arc.
%
%

In all simulations, the weights are chosen as $w_v = 1$, $w_X = w_Y = 10$ and $w_o = 100$. The planning horizon is chosen as $T = \SI{3}{\second}$ and the time step duration of the MPC is $h = \SI{200}{\milli\second}$; replanning is performed every \SI{100}{\milli\second}. To achieve real-time computation speeds, the solver is limited to five SQP iterations, which in our experience is sufficient to achieve a good convergence. Auto-generation techniques can also be used to further reduce computation time~\cite{Frasch2013}.

For comparison purposes, we also implemented the same MPC planner with a classical kinematic bicycle model such as presented in~\cite{Kong2015}; the model is written as follows:
\begin{subequations}
\begin{align}
\dot X &\ = v \cos\left(\psi + \arctan \beta \right) \\
\dot Y &\ = v \sin\left(\psi + \arctan \beta \right) \\
\dot \psi &\ = \frac{v}{l_r} \sin\left(\arctan \beta \right) \\
\dot v &\ = a
\end{align}
\end{subequations}
with $v$ the longitudinal velocity and $\beta = \frac{l_r}{l_r+l_f} \tan \delta$ the side slip angle. The control inputs are the longitudinal acceleration $a$, and the steering angle of the front wheel $\delta$. 

All other parts of the planning and control algorithm are otherwise equal, including the low-level controller. The maximum and minimum acceleration in the bicycle model are chosen equal to $a_X^{min}$ and $a_X^{max}$ (see \Cref{fig:polygon-approx} for the notations) respectively. Note that the solver is slightly faster using this formulation; therefore, the maximum number of SQP iterations is set to 6 for the kinematic model to yield comparable computation times, which increases solution quality.

\subsection{Planning without obstacles}
\begin{figure}
	\includegtikz[width=0.9\columnwidth,height=4cm]{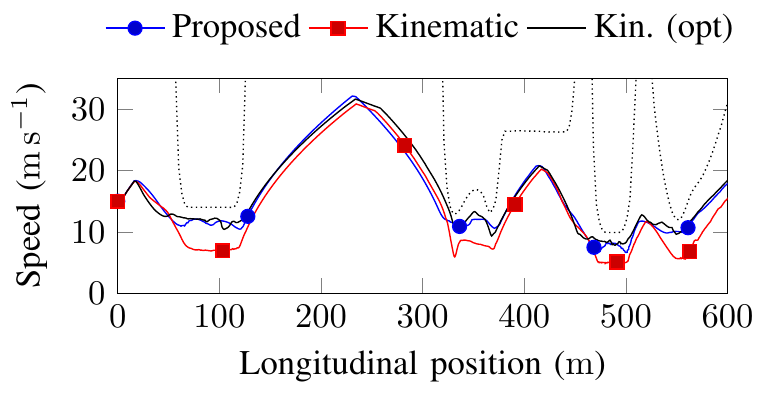}
	\caption{Comparison of achieved vehicle speed with our proposed model (in blue) and for a kinematic bicycle model (in red); the speed for the kinematic model run until convergence is shown in solid black. The dotted curve represents the bound $v^2 \leq \mu g / \kappa$ where $\kappa$ is the path curvature. \label{fig:compspeed-noobst}}
\end{figure}

We first consider trajectory planning without obstacles; \Cref{fig:compspeed-noobst} presents the actual speed of the vehicle during the simulation as a function of its position along the path $\gamma_{ref}$ for the two MPC planners, as well as the speed bound $v^2 \leq \kappa g$ (with $\kappa$ the path curvature), corresponding to a centripetal acceleration of $1\, g$ used, \textit{e.g.}, in~\cite{koh2016speed}. First, we notice that both planners have similar performance in the straight portions of the road; however, the planner based on our proposed second-order integrator model systematically achieves higher speeds in curves. Second, no solver reaches the upper bound $\sqrt{g/\kappa}$, thus confirming that including a speed selection phase during planning (as opposed to tracking a predefined velocity solely computed from path curvature) is relevant to allow proper tracking.

The superior performance of the planner based on the second-order integrator model is likely due to the simpler relation between the optimization variables (the input controls) and the objective value (the deviation from the target state) than in the bicycle model, which allows a much faster convergence towards the solution. Indeed, when allowing the solver to run until convergence with the kinematic bicycle model, the resulting velocity becomes comparable to that obtained with the real-time second-order model (but computation time is above \SI{500}{\milli\second}).

\begin{figure}
	\includegtikz[width=0.9\columnwidth,height=4cm]{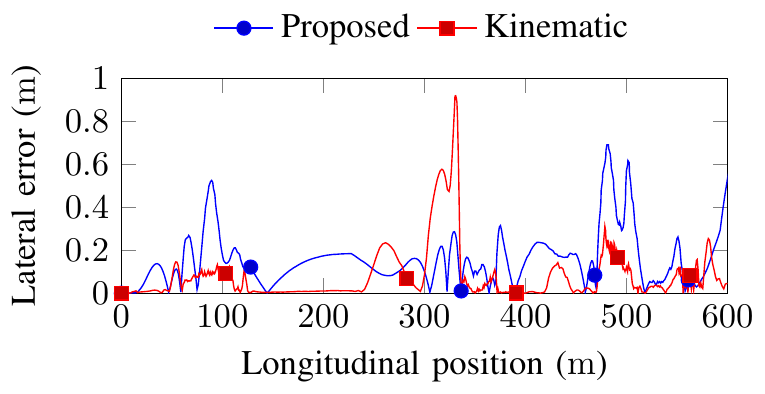}
	\caption{Lateral positioning error for the proposed model (in blue) and for the kinematic bicycle model (in red), without obstacles. \label{fig:comperr-noobst}}
\end{figure}

\Cref{fig:comperr-noobst} shows the lateral error when tracking the reference path, for both MPC planners. \Cref{tab:comp-error} presents synthetic data about the lateral error of the complete planning and control architecture in both cases, showing satisfying overall performance for high-speed applications. Note that better precision can be achieved (at the cost of speed) by selecting different values for the weighting coefficients. Moreover, a more precise low-level controller can probably achieve better performance.

\begin{table}\centering
	\caption{Absolute lateral positioning error for both planners.\label{tab:comp-error}}
	\begin{tabular}{l c c}
		\toprule 
		Model & RMS error (\SI{}{\meter}) & Maximum error (\SI{}{\meter}) \\
		\midrule 
		Proposed & $0.25$ & $0.70$ \\
		Kinematic & $0.16$ & $0.95$ \\
		\bottomrule
	\end{tabular}
\end{table}

\subsection{Planning with obstacle avoidance}
In this section, we compare the behavior of both planners in the presence of obstacles, modeled as parabolas as explained in \Cref{fig:parabola}. \Cref{fig:output-obst} shows a detail of the actual path followed by the vehicle using the proposed planner while avoiding obstacles. The attached video file\footnote{Also available at \url{https://youtu.be/BRpmdIxTz-0}} shows the corresponding simulation.


\begin{figure}
	\centering \includegtikz[width=0.9\columnwidth,height=4cm]{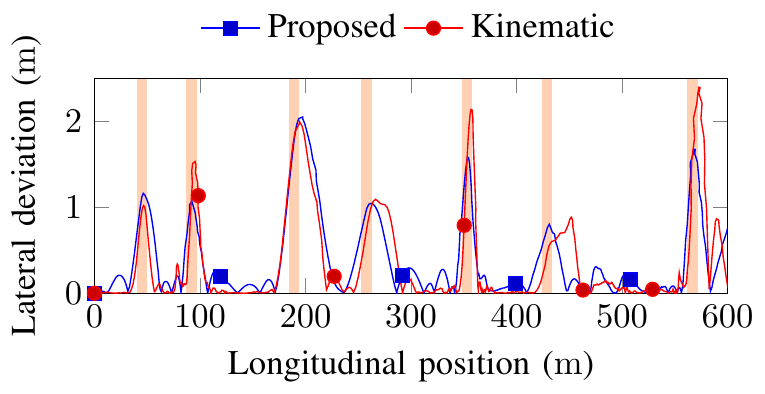}
	\caption{Lateral deviation for the proposed model (in blue) and for the kinematic bicycle model (in red) while avoiding obstacles (shaded regions). \label{fig:comperr-obst}}
\end{figure}

\begin{figure}
	\includegtikz[width=0.9\columnwidth,height=3.5cm]{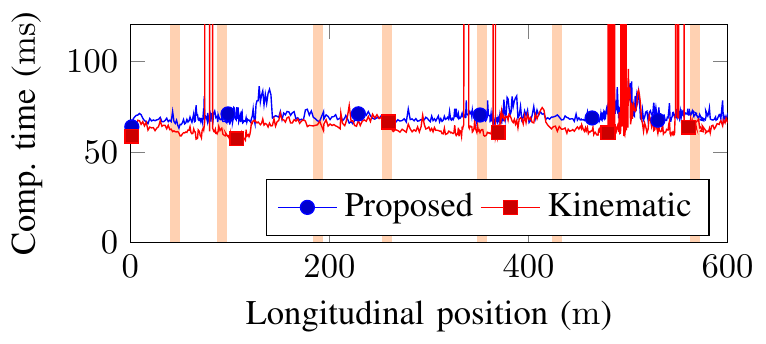}
	\caption{MPC computation time (with obstacles) for the proposed model, and for a kinematic bicycle model starting from the same state. \label{fig:comput-time}}
\end{figure}

\Cref{fig:comperr-obst} presents the lateral deviation from the reference trajectory while avoiding obstacles using both planners. Generally speaking, the proposed planner allows a smaller deviation from the reference and a higher average speed of \SI{12.7}{\meter\per\second} compared to \SI{10.2}{\meter\per\second} using the kinematic bicycle planner (for lack of space, the velocity curves are not shown). More importantly, the kinematic planner is sometimes unable to output a trajectory in less than \SI{100}{\milli\second}, as shown in \Cref{fig:comput-time}. This situation happens when approaching obstacles in high-curvature portions of the road; as before, the better behavior of the proposed planner is likely due to the simpler search space since the dynamic model presents much less non-linearity. A more in-depth analysis could provide useful insights on desirable model properties to allow fast and robust convergence.

\section{Conclusion\label{sec:conclusion}}
In this article, we considered the trajectory planning and control of a vehicle at high velocity near the limits of handling. Instead of using a highly complex model during online resolution, as it is often the case in the literature, we used a simpler vehicle model obtained from precise dynamic simulations. This model was implemented into a novel MPC-based trajectory planner; contrary to most existing algorithms, our formulation does not need a predefined target speed. Instead, the MPC adjusts the vehicle's speed in real time to track a predefined path (such as the centerline of a road) as fast as possible while maintaining a low tracking error. Using the high fidelity physics simulator PreScan, we demonstrated that the combination of this planner with a simple, PID-based low-level controller is capable of driving along a demanding path at high speed with a low lateral error. Moreover, comparison with a simple and widely used bicycle model shows that the use of our simpler model allows the solver to converge faster towards a better quality solution under real-time constraints.

Although this work remains mainly theoretical, it opens several perspectives for future research. First, the good performance of our simple dynamic model even at high speeds allows envisioning longer planning horizons without sacrificing computation speed. Future work should also study the consistency of using fully linear models, that can be coupled with efficient mixed-integer optimization techniques to allow optimal decision-making, for instance for overtaking or lane-change decisions~\cite{Qian2016}. Second, we believe that the ability of the proposed MPC formulation to automatically adapt the vehicle's velocity can find practical applications, for instance when driving in low-adherence situations. Therefore, the behavior of the planner with a variable friction coefficient should be studied further. A more precise low-level controller, taking into account the very particular response curve of the tires to precisely track a target force, which would be more consistent with the second order integrator model, can also prove interesting. Finally, implementation of the planner and controller on an actual (scale model) vehicle is a necessary step to validate the suitability of the proposed planner for real-world use.

\bibliographystyle{IEEEtranurl}
\bibliography{dynamic}

\begin{thebibliography}{10}
\providecommand{\url}[1]{#1}
\csname url@rmstyle\endcsname
\providecommand{\newblock}{\relax}
\providecommand{\bibinfo}[2]{#2}
\providecommand\BIBentrySTDinterwordspacing{\spaceskip=0pt\relax}
\providecommand\BIBentryALTinterwordstretchfactor{4}
\providecommand\BIBentryALTinterwordspacing{\spaceskip=\fontdimen2\font plus
\BIBentryALTinterwordstretchfactor\fontdimen3\font minus
  \fontdimen4\font\relax}
\providecommand\BIBforeignlanguage[2]{{%
\expandafter\ifx\csname l@#1\endcsname\relax
\typeout{** WARNING: IEEEtran.bst: No hyphenation pattern has been}%
\typeout{** loaded for the language `#1'. Using the pattern for}%
\typeout{** the default language instead.}%
\else
\language=\csname l@#1\endcsname
\fi
#2}}

\bibitem{Benenson2008}
\BIBentryALTinterwordspacing
R.~Benenson, S.~Petti, T.~Fraichard, and M.~Parent, ``{Towards urban driverless
  vehicles},'' \emph{International Journal of Vehicle Autonomous Systems},
  vol.~6, no. 1/2, p.~4, 2008.
\BIBentrySTDinterwordspacing

\bibitem{Qian2016}
\BIBentryALTinterwordspacing
X.~Qian, F.~Altch\'e, P.~Bender, C.~Stiller, and A.~{{de} La Fortelle},
  ``{Optimal trajectory planning for autonomous driving integrating logical
  constraints: An MIQP perspective},'' in \emph{2016 IEEE 19th International
  Conference on Intelligent Transportation Systems (ITSC)}.\hskip 1em plus
  0.5em minus 0.4em\relax IEEE, nov 2016, pp. 205--210.
\BIBentrySTDinterwordspacing

\bibitem{Falcone2008}
\BIBentryALTinterwordspacing
P.~Falcone, F.~Borrelli, H.~E. Tseng, J.~Asgari, and D.~Hrovat, ``{A
  hierarchical Model Predictive Control framework for autonomous ground
  vehicles},'' \emph{2008 American Control Conference}, pp. 3719--3724, 2008.
\BIBentrySTDinterwordspacing

\bibitem{Frasch2013}
J.~Frasch, A.~Gray, M.~Zanon, H.~Ferreau, S.~Sager, F.~Borrelli, and M.~Diehl,
  ``{An auto-generated nonlinear MPC algorithm for real-time obstacle avoidance
  of ground vehicles},'' \emph{European Control Conference (ECC), 2013}, pp.
  4136--4141, 2013.

\bibitem{koh2016speed}
Y.~Koh, K.~Yi, H.~Her, and K.~Kim, ``A speed control race driver model with
  on-line driving trajectory planning,'' in \emph{The Dynamics of Vehicles on
  Roads and Tracks: Proceedings of the 24th Symposium of the International
  Association for Vehicle System Dynamics (IAVSD 2015), Graz, Austria, 17-21
  August 2015}.\hskip 1em plus 0.5em minus 0.4em\relax CRC Press, 2016, p.~67.

\bibitem{Kim2011}
\BIBentryALTinterwordspacing
D.~Kim, J.~Kang, and K.~Yi, ``{Control strategy for high-speed autonomous
  driving in structured road},'' in \emph{2011 14th International IEEE
  Conference on Intelligent Transportation Systems (ITSC)}.\hskip 1em plus
  0.5em minus 0.4em\relax IEEE, oct 2011, pp. 186--191.
\BIBentrySTDinterwordspacing

\bibitem{Pacejka2005}
H.~Pacejka, \emph{Tire and vehicle dynamics}.\hskip 1em plus 0.5em minus
  0.4em\relax Elsevier, 2005.

\bibitem{Altrock1994}
C.~V. Altrock, ``Fuzzy logic technologies in automotive engineering,'' in
  \emph{WESCON/94. Idea/Microelectronics. Conference Record}, Sep 1994, pp.
  110--117.

\bibitem{Abbas2014}
\BIBentryALTinterwordspacing
M.~A. Abbas, R.~Milman, and J.~M. Eklund, ``{Obstacle avoidance in real time
  with Nonlinear Model Predictive Control of autonomous vehicles},'' in
  \emph{2014 IEEE 27th Canadian Conference on Electrical and Computer
  Engineering (CCECE)}.\hskip 1em plus 0.5em minus 0.4em\relax IEEE, may 2014,
  pp. 1--6.
\BIBentrySTDinterwordspacing

\bibitem{Cardoso2016}
\BIBentryALTinterwordspacing
V.~Cardoso, J.~Oliveira, T.~Teixeira, C.~Badue, F.~Mutz, T.~Oliveira-Santos,
  L.~Veronese, and A.~F. {De Souza}, ``{A Model-Predictive Motion Planner for
  the IARA Autonomous Car},'' \emph{arXiv preprint arXiv:1611.04552}, nov 2016.
\BIBentrySTDinterwordspacing

\bibitem{Falcone2007}
\BIBentryALTinterwordspacing
P.~Falcone, F.~Borrelli, J.~Asgari, H.~E. Tseng, and D.~Hrovat, ``{Predictive
  Active Steering Control for Autonomous Vehicle Systems},'' \emph{IEEE
  Transactions on Control Systems Technology}, vol.~15, no.~3, pp. 566--580,
  may 2007.
\BIBentrySTDinterwordspacing

\bibitem{Park2009}
\BIBentryALTinterwordspacing
J.-M. Park, D.-W. Kim, Y.-S. Yoon, H.~J. Kim, and K.-S. Yi, ``{Obstacle
  avoidance of autonomous vehicles based on model predictive control},''
  \emph{Proc. of the Institution of Mechanical Engineers, Part D: Journal of
  Automobile Engineering}, vol. 223, no.~12, pp. 1499--1516, 2009.
\BIBentrySTDinterwordspacing

\bibitem{Gao2010}
\BIBentryALTinterwordspacing
Y.~Gao, T.~Lin, F.~Borrelli, E.~Tseng, and D.~Hrovat, ``{Predictive Control of
  Autonomous Ground Vehicles With Obstacle Avoidance on Slippery Roads},'' in
  \emph{ASME 2010 Dynamic Systems and Control Conference, Volume 1}.\hskip 1em
  plus 0.5em minus 0.4em\relax ASME, 2010, pp. 265--272.
\BIBentrySTDinterwordspacing

\bibitem{liniger2015optimization}
A.~Liniger, A.~Domahidi, and M.~Morari, ``Optimization-based autonomous racing
  of 1: 43 scale rc cars,'' \emph{Optimal Control Applications and Methods},
  vol.~36, no.~5, pp. 628--647, 2015.

\bibitem{Liu2014}
\BIBentryALTinterwordspacing
J.~Liu, P.~Jayakumar, J.~L. Stein, and T.~Ersal, ``{A Multi-Stage Optimization
  Formulation for MPC-Based Obstacle Avoidance in Autonomous Vehicles Using a
  LIDAR Sensor},'' in \emph{ASME 2014 Dynamic Systems and Control
  Conference}.\hskip 1em plus 0.5em minus 0.4em\relax ASME, oct 2014.
\BIBentrySTDinterwordspacing

\bibitem{JeonghwanJeon2013}
\BIBentryALTinterwordspacing
J.~Jeon, R.~V. Cowlagi, S.~C. Peters, S.~Karaman, E.~Frazzoli, P.~Tsiotras, and
  K.~Iagnemma, ``{Optimal motion planning with the half-car dynamical model for
  autonomous high-speed driving},'' in \emph{2013 American Control
  Conference}.\hskip 1em plus 0.5em minus 0.4em\relax IEEE, jun 2013, pp.
  188--193.
\BIBentrySTDinterwordspacing

\bibitem{Altche2017}
\BIBentryALTinterwordspacing
F.~Altch\'e, P.~Polack, and A.~{{de} La Fortelle}, ``{A Simple Dynamic Model
  for Aggressive, Near-Limits Trajectory Planning},'' \emph{arXiv preprint
  arXiv:1703.01225}, mar 2017.
\BIBentrySTDinterwordspacing

\bibitem{guntur1980friction}
R.~Guntur and S.~Sankar, ``A friction circle concept for dugoff's tyre friction
  model,'' \emph{International Journal of Vehicle Design}, vol.~1, no.~4, pp.
  373--377, 1980.

\bibitem{Kong2015}
\BIBentryALTinterwordspacing
J.~Kong, M.~Pfeiffer, G.~Schildbach, and F.~Borrelli, ``{Kinematic and dynamic
  vehicle models for autonomous driving control design},'' in \emph{2015 IEEE
  Intelligent Vehicles Symposium (IV)}.\hskip 1em plus 0.5em minus 0.4em\relax
  IEEE, jun 2015, pp. 1094--1099.
\BIBentrySTDinterwordspacing

\bibitem{qian2016hierarchical}
X.~Qian, A.~De~La~Fortelle, and F.~Moutarde, ``A hierarchical model predictive
  control framework for on-road formation control of autonomous vehicles,'' in
  \emph{Intelligent Vehicles Symposium (IV), 2016 IEEE}.\hskip 1em plus 0.5em
  minus 0.4em\relax IEEE, 2016, pp. 376--381.

\bibitem{Gray2013}
\BIBentryALTinterwordspacing
A.~Gray, Y.~Gao, J.~K. Hedrick, and F.~Borrelli, ``{Robust Predictive Control
  for semi-autonomous vehicles with an uncertain driver model},'' in \emph{2013
  IEEE Intelligent Vehicles Symposium (IV)}.\hskip 1em plus 0.5em minus
  0.4em\relax IEEE, jun 2013, pp. 208--213.
\BIBentrySTDinterwordspacing

\bibitem{Borrelli2005}
\BIBentryALTinterwordspacing
F.~Borrelli, P.~Falcone, T.~Keviczky, J.~Asgari, and D.~Hrovat, ``{MPC-based
  approach to active steering for autonomous vehicle systems},''
  \emph{International Journal of Vehicle Autonomous Systems}, vol.~3, no.
  2/3/4, p. 265, 2005.
\BIBentrySTDinterwordspacing

\bibitem{prescan}
\BIBentryALTinterwordspacing
{TASS International}. \url{http://www.tassinternational.com/prescan}.
\BIBentrySTDinterwordspacing

\bibitem{Houska2011}
B.~Houska, H.~Ferreau, and M.~Diehl, ``{ACADO} {T}oolkit -- {A}n {O}pen
  {S}ource {F}ramework for {A}utomatic {C}ontrol and {D}ynamic
  {O}ptimization,'' \emph{Optimal Control Applications and Methods}, vol.~32,
  no.~3, pp. 298--312, 2011.

\bibitem{Kosecka1997}
\BIBentryALTinterwordspacing
J.~Kosecka, R.~Blasi, C.~Taylor, and J.~Malik, ``{Vision-based lateral control
  of vehicles},'' in \emph{Proceedings of Conference on Intelligent
  Transportation Systems}, vol.~46.\hskip 1em plus 0.5em minus 0.4em\relax
  IEEE, 1997, pp. 900--905.
\BIBentrySTDinterwordspacing

\bibitem{breuer1998}
J.~Breuer, ``Analysis of driver-vehicle-interactions in an evasive manoeuvre -
  results of ``moose test'' studies,'' \emph{Proceedings: International
  Technical Conference on the Enhanced Safety of Vehicles}, vol. 1998, pp.
  620--627, 1998.

\end{thebibliography}

\end{document}